# A Novel Approach with Monte-Carlo Simulation and Hybrid Optimization Approach for Inventory Management with Stochastic Demand

Sarit Maitra
Alliance Business School
Alliance University,
Bengaluru, India
sarit.maitra@alliance.edu.in

Vivel Mishra
School of Applied Mathematics
Alliance University,
Bengaluru, India
vivek.mishra@alliance.edu.in

Sukanya Kundu
Alliance Business School
Alliance University,
Bengaluru, India
sukanya.kundu@alliance.edu.in

**Abstract**

This study addresses the difficulties associated with inventory management of products with stochastic demand. The objective is to find the optimal combination of order quantity and reorder point that maximizes profit while considering ethical considerations in inventory management. The ethical considerations are risk assessment, social responsibility, environmental sustainability, and customer satisfaction. Monte Carlo simulation (MCS) is used in this study to generate a distribution of demand and lead times for the inventory items, which is then used to estimate the potential profit and risk associated with different inventory policies. This work proposes a hybrid optimization approach combining Gaussian process regression and conditioning function to efficiently search the high- dimensional space of potential continuous review (r, Q) and periodic review (p, Q) values to find the optimal combination that maximizes profit while considering ethical considerations. The findings show that both the (r, Q) and (p, Q) approaches can effectively manage inventory with stochastic demand, but the (r, Q) approach performs better (profits up by 12.73%) when demand is more volatile. The study adds quantifiable risk assessment and sensitivity analysis to these considerations, considering the variation in demand and expected output in profit percentage. The results provide useful information for making ethical and responsible choices in supply chain analytics, boosting efficiency and profits.

1. Introduction

According to the 34th Annual Council of Supply Chain Management Professionals State of Logistics report, U.S. business logistics costs stands at a record $2.3 trillion (around 9% of US GDP). Inventory management (IM) stands as a cornerstone in supply chain management, wielding significant influence over a company's financial performance. The intricacies of IM are accentuated when confronted with stochastic demand, which can lead to consequences such as stockouts, excessive inventory holdings, revenue losses, etc. The primary objective of inventory management is to optimize the ordering quantity and reorder point for products subject to stochastic demand. The main goal is to maximize profitability of an organization while simultaneously minimizing overall costs by reducing stockouts and revenue losses. Keeping this in mind, the purpose of this research is to find a strategy for dealing with stochastic demand.

Although a modern supply chain is too complex for an equation to describe, we can nevertheless attempt to simulate it. When no equations are up to the task, simulations can be used to either establish a good enough policy or to test the results of a mathematical model (Vandeput, 2020). The effectiveness of simulation modeling has been highlighted by several researchers (e.g., Costantino et al., 2014; Lowalekar & Ravichandran, 2014; Logachev et al., 2022, etc.). This study uses Monte Carlo simulation (MCS) to generate demand and lead time distributions for inventory items, estimating potential profits and risks under different inventory policies. MCS provides insights and optimization in complex, uncertain environments, making it a practical approach in operations research and supply chain management (Slama et al., 2021; Weraikat et al., 2019; Patriarca et al., 2020; Harifi et al., 2021). It employs Gaussian process regression (GPR) to predict demand and Bayesian optimization to account for lead time and order costs. GPR is effective at anticipating demand and improving decision-making in uncertain situations since it can capture intricate and non-linear patterns in inventory data while offering probabilistic predictions. Gaussian process regression is a non-parametric regression method that uses conditioning on Gaussian vectors to find a model that passes through the data points. It requires specifying covariance functions and optimizing hyperparameter values. Due to its probabilistic nature, the classic model optimization approach is not suitable, requiring a probabilistic approach like the maximum likelihood method (Ahmad et al., 2022).

The findings provide valuable insights for responsible inventory management decisions. Additionally, this study introduces conditioning functions. It allows the algorithm to adapt and enhance its knowledge of the objective function as additional data is gathered, assisting in the search for optimal solutions in high-dimensional and uncertain parameter spaces.

Moreover, the effectiveness of inventory policies significantly hinges on the underlying demand distribution and the variability in lead times. This inherently introduces elements of risk and uncertainty into managing inventory. Muller (2019) and Liu et al. (2022) contributed significantly to inventory risk management by emphasizing the importance of risk assessment and its profound impact on inventory management strategies. To address the risk factor, sensitivity analysis takes center stage and is an important aspect of this research. Sensitivity analysis allows us to assess the robustness of the inventory policy when subjected to diverse and volatile scenarios. This involves evaluating the efficacy of the policy considering multiple performance metrics, including stockout rate, average inventory levels, and total cost.

In essence, this research aims to improve inventory management decisions in uncertain environments. It uses an analytical and simulation-based framework that promises a deeper understanding of inventory policies and optimization techniques. Inventory management being one of the major cost components of supply chain, the study aligns with the Supply Chain Analytics Journal's objectives, enhancing businesses' competitiveness and relevance in uncertain eras.

## 2. Literature review

This literature review explores inventory management (IM) in modern supply chains, highlighting the shift from deterministic to stochastic demand modeling, the development of analytical and systematic approaches, and the significance of optimization techniques in addressing contemporary IM complexities.

Historically, inventory management has majorly focused on deterministic demand models (e.g., Babiloni & Guijarro, 2020; Stopková et al., 2019; Tamjidzad & Mirmohammadi, 2017 & 2015; Choi, 2013; etc.). However, the applicability of such models in real-world scenarios has been increasingly limited due to long list of constraints and the rising unpredictability and uncertainty in demand patterns. Recent researchers have turned their focus to stochastic demand modeling, which provides a flexible and principled method for probabilistic modelling and inference. (e.g., Heikkinen et al., 2022; Li et al., 2021; Wang et al., 2020; Gholami et al., 2021; Pearce et al., 2022). Stochastic demand and probabilistic models are interconnected because stochastic demand inherently involves uncertainty and variability, which are core concepts addressed by probabilistic modeling (Hasan et al., 2019; Sakki et al., 2022). Moreover, with the probabilistic model, researchers have started investigating simulation modeling (e.g., Sridhar et al., 2021; Becerra et al., 2021; Evans & Bae, 2019). Syntetos et al. (2016) and Gonçalves et al. (2020) have leveraged machine learning algorithms to forecast demand variability and propose effective safety stock calculation methods, further advancing inventory management practices. In line with recent advancements, Ekren et al. (2023) introduced a demand forecasting model that not only considers demand variability but also employs a robust optimization approach, aligning with the ongoing efforts to enhance inventory management strategies. The evolving nature of supply chains necessitated sophisticated inventory management approaches, leveraging the power of stochastic demand modeling, probabilistic methods, simulation modeling, machine learning, and robust optimization techniques.

The significance of adopting data-driven, analytical, and systematic approaches is imperative for effectively tackling challenges in inventory management (Anderson et al., 2018). While several studies (e.g., Maddah and Noueihed, 2017; Taleizadeh et al., 2020; Zeng et al., 2019, among others) delved into the economic order quantity (EOQ) formula, they lacked rigorous analytical evaluations to substantiate their findings. Notably, the relevance of the (r, Q) reorder point and order quantity policy in enhancing service levels and reducing inventory costs amid stochastic demand has been convincingly demonstrated in the works of De & Sana (2018), Das et al. (2019), Braglia et al. (2019), and Wang et al. (2022). Extending this concept further, Tang et al. (2018) incorporated both cost and emissions considerations, aligning with the sustainability goals of modern supply chain management. This emphasizes the macro level perspective and even highlights the dynamic nature of inventory management research and the importance of in-depth multiple parameter study and data-driven decision-making. It also emphasizes how well (r, Q) strategies work to solve problems brought on by stochastic demand.

Optimization techniques play an indispensable role in implementing appropriate inventory policies (Wu & Frazier, 2019; Gruler et al., 2018; Azadi et al., 2019). To discover nearly optimal control parameters for a stochastic multi-product inventory control system, Jackson (2019) suggested a simulation-optimization framework, allowing for full inventory dynamics with risk and reliability analysis. However, these techniques may not guarantee optimal solutions if underlying assumptions are violated (Jackson et al. 2020). The underlying assumptions include deterministic demand, fixed lead times, constant costs, independent and identically distributed assumption, continuous review, linear relationships, and known probability distributions. Qiu et al. (2021) emphasizes the need to account for uncertainty and variability in inventory management models to address potential disruption risks. Supply chain risks are numerous and can be divided into operational and disruptive risks (Choi et al., 2019; Xu et al., 2020). Operational risks involve daily disturbances in inventory operations like lead-time and demand fluctuations, while disruption risks involve low-frequency, high-impact events (Kinra et al., 2020; Hosseini & Ivanov, 2022). The recent pandemic has significantly disrupted the supply chain, leading to material shortages, delivery delays, and performance degradation, affecting revenue, service levels, and productivity; all these come under disruption risk (Dolgui, 2020; Pengfei et al., 2021; Li & Zobel, 2020). Ivanov (2020) has highlighted the importance of simulation-based models. They found that these models are effective in observing and predicting SC behaviors over time, adding dynamic features to optimization techniques commonly used in SC risk analysis. In their subsequent work, they highlighted that simulation models consider logical and randomness constraints, including disruptions, inventory, production, sourcing, and shipment control policies, and gradual capacity degradation and recovery (Ivanov & Dolgui, 2021). Sharma et al. (2020) highlighted the importance of balancing risk and operational flexibility in supply network design, managing supply chain disruptions, and maintaining agility in the face of demand uncertainty. All the above studies collectively highlight the importance of optimization techniques and simulation-based models in modern inventory management, enhancing efficiency, resilience, and risk mitigation in supply chain complexities. It is practically impossible to optimize inventory policy with 100% accuracy in terms of model accuracy and optimization (Venderput, 2022). Any supply chain can benefit from significant cost reductions by having information that is accurate 95% of the time.

While the existing literature has made substantial strides in addressing the challenges of stochastic demand and optimizing inventory policies, it is imperative to acknowledge that the field of supply chain analytics continues to evolve rapidly. Researchers (e.g., Perez et al., 2021) worked on inventory management policies in a supply network and found that stochastic linear programming is more profitable, while reinforcement learning is more robust to disruptions, however, they suggested potential areas for further research in non-stationary demand and end-of-simulation effects. Our findings may pave the way for future investigations into these areas, further enriching the field of inventory management. We believe that our integrated optimization approach opens exciting avenues for future research in the dynamic inventory management landscape. By demonstrating the effectiveness of this approach, we hope to encourage further exploration of its applications and refinements.

## 3. Methodology

We propose statistical analysis, simulation modeling using Monte Carlo simulations, and Bayesian Optimization with a conditioning function to compare inventory management policies under stochastic demand. For empirical analysis, we collected sales data from multiple products to model demand. Multiple scenarios were simulated 10,000 times to determine optimal inventory levels. The simulation runs for 365 days (about 12 months) to optimize savings and profits.

### 3.1. Uncertainty

The major sources of uncertainty identified here are:

1. Unpredictable purchase behavior of customer:

p = (*number of orders last year*)/(*number of working days*), where p is the likelihood of a customer placing an order on any given day. However, this model is not perfect, and there may be some level of error in the estimated likelihood of a customer placing an order.
2. Variability in order size is modelled using a lognormal distribution and represents a second cause of uncertainty. While past sales data is used to predict distribution parameters, actual order sizes may still differ from these projections.

We have taken steps to address these uncertainties by using Poisson and lognormal distributions in modeling and simulation. These distributions account for variability and randomness in both customer purchase behavior and order sizes, making the analysis more robust and adaptable to real-world fluctuations. Besides, two inventory management policies are considered in this study: (r, Q): reorder quantity continuous review; and (p, Q): periodic quantity review. The performance of these policies is compared based on their ability to minimize inventory holding costs while ensuring an acceptable level of service. To evaluate the performance, the study uses various performance metrics, such as expected inventory holding costs, expected backorder costs, and expected fill rate. The simulation results are then analyzed and interpreted to provide insights into each policy's effectiveness.

### 3.2    Simulation-Based Algorithm Optimization

Bayesian optimization (BO) leverages data-driven techniques and surrogate modeling to efficiently optimize inventory policies. It balances exploration and exploitation for global optimization and is well-suited for efficiently maximizing service levels and profit. Fig. 1 depicts the fundamental concepts of Bayesian optimization utilizing GP, as well as the algorithm and how it works.

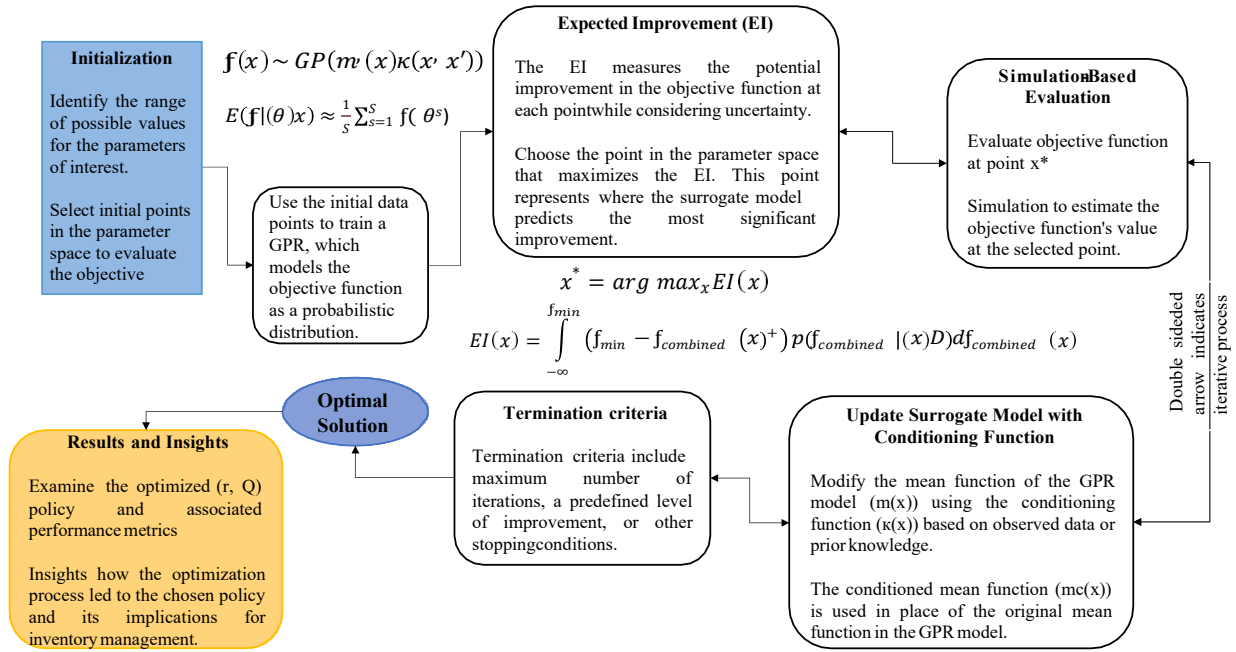

Fig. 1    Workflow for Bayesian Optimization with Conditioning Function

The Bayesian model is solved by using simulation methods. For a function $f$ depending on $\theta$, a vector of unknown parameters to be estimated based on data $x$,

$$E(f(\theta)|x) \approx \frac{1}{S}\sum_{s=1}^{S} f(\theta^s) \qquad (1)$$

Where, $\theta^s$ represents a simulated sample from the posterior probability distribution of and is the total number of simulated samples. BO uses the objective function of Gaussian Process Regression (GPR) and the expected improvement acquisition function. It quantifies the objective function of GPR by evaluating a potential $x$. The algorithm builds a surrogate model of GPR and uses it to select the next point to evaluate. The $EI$ acquisition function balances exploration and exploitation to choose the next point that is most likely to improve the surrogate model's performance.

The algorithm selects the next point for evaluation, denoted as $x^*$, by maximizing the EI:

$$x^* = arg\ max_x EI(x) \qquad (2)$$

The $EI$ acquisition function balances exploration and exploitation by considering both the predicted performance improvement at a point and the uncertainty associated with that prediction. This balance allows the algorithm to choose the next evaluation point $x^*$ that enhances the surrogate model's performance.

4. **Data Analysis**

The business case selected here examines the sale of four distinct products. The goal is to maximize the anticipated profit. Table 1 displays the descriptive statistics computed based on historical demand data (365 days). Some general inferences are made from the given dataset as stated below.

Table 1. Product data

|  | PrA | PrB | PrC | PrD |
|---|---|---|---|---|
| PurchaseCost | € 12 | € 7 | € 6 | € 37 |
| LeadTime | 9 | 6 | 15 | 12 |
| Size | 0.57 | 0.05 | 0.53 | 1.05 |
| SellingPrice | € 16.10 | € 8.60 | € 10.20 | € 68 |
| StartingStock | 2750 | 22500 | 5200 | 1400 |
| Mean | 103.50 | 648.55 | 201.68 | 150.06 |
| StdDev | 37.32 | 26.45 | 31.08 | 3.21 |
| OrderCost | € 1000 | € 1200 | € 1000 | € 1200 |
| HoldingCost | € 20 | € 20 | € 20 | € 20 |
| Probability | 0.76 | 1.00 | 0.70 | 0.23 |
| DemandLead | 705 | 3891 | 2266 | 785 |
| Annual demand | 28,670 | 237,370 | 51,831 | 13,056 |

- Purchase Cost: PrD has the highest purchase cost, indicating it might be a higher-priced or specialized product. PrA, PrB, and PrC have relatively lower purchase costs, suggesting they are less expensive items.
- Lead Time: PrC has the highest lead time, which means it takes the longest to replenish inventory for this product. PrB has the shortest lead time, implying it can be restocked relatively quickly.
- Size: PrD has the largest size among the products, possibly indicating it requires more storage space. PrB has the smallest size, which could make it easier to manage in terms of storage.

- Selling Price: PrD has the highest selling price, which may indicate it's a high-value product with potentially higher profit margins. PrA and PrC have moderate selling prices. PrB has the lowest selling price, suggesting it may be a lower-cost item.
- Starting Stock: PrB has significantly higher starting stock compared to other products, indicating it might be in higher demand or part of a different inventory strategy.
- Mean Demand: PrB has the highest mean demand, suggesting it's the most popular product. PrD has the lowest mean demand, indicating it may be a niche or specialty item.
- Standard Deviation of Demand (StandardDev): PrD has the lowest standard deviation of demand, indicating it has more consistent demand over time. PrA and PrC have relatively higher demand variability.
- Order Cost and Holding Cost: All the products have the same order and holding costs, which simplifies the cost structure for inventory management.
- Probability: PrB has a probability of 1.00, suggesting it always needs to be in stock. PrD has the lowest probability, indicating it's ordered less frequently, possibly due to lower demand or a higher purchase cost.
- Demand Lead Time: PrB has the highest demand lead time, which could be due to its larger starting stock and higher demand.

Fig. 2 displays the distribution of each product.

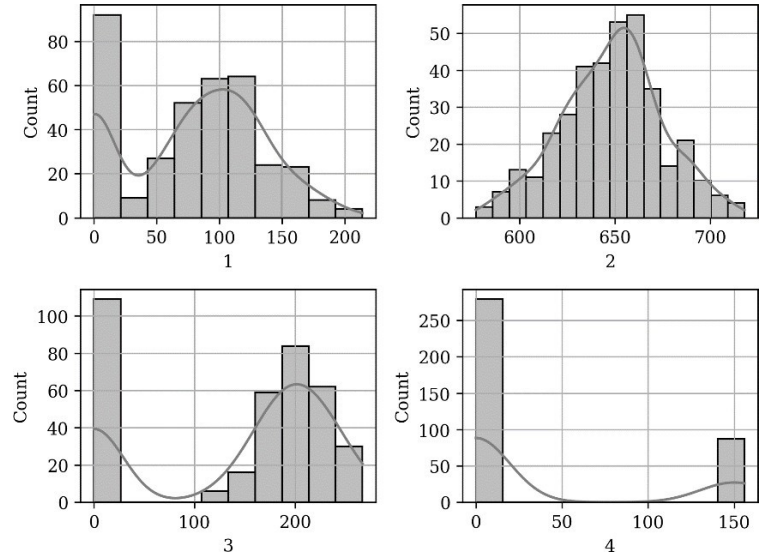

Fig. 2. Distribution of Product Demand as a Histogram

Based on the data in Table 1, the EOQ model produces a report detailing the annual cost, optimal order quantity, annual profit, anticipated proportion of lost orders, lead time variability, safety stock, and reorder point (ROP).

- $EOQ = \sqrt{2 * AnnualDemand * OrderCost/HoldingCost}$
- $Total\ Annual\ Cost = (Annual\ Demand\ /\ EOQ) * Order\ Cost + (EOQ\ /\ 2) * Holding\ Cost$
- $Total\ Annual\ Profit = (Annual\ Demand * Selling\ Price) - Total\ Annual\ Cost$
- $Expected\ Proportion\ of\ Lost\ Order = Probability\ of\ Stockout * (1 - Safety\ Stock\ /\ Average\ Inventory)$
- $Safety\ Stock = Zscore * StdDev\ of\ Demand * \sqrt{leadTime + ReviewTime}$
- $Reorder\ Point\ (ROP) = Lead\ Time\ Demand + Safety\ Stock$

Above are standard formulas used in inventory management, specifically for the Economic Order Quantity (EOQ) model. These formulas are widely recognized and accepted in the fields of operations management and inventory control. Table 2 presents a report for each product, assuming a service level of 95% and a Z-score of 1.65 (corresponding to 95% of the standard normal distribution).

$$\text{EOQ (Economic Order Quantity)} = \sqrt{2 * Annual\ Demand * Order\ Cost) / Holding\ Cost\ per\ Unit}$$

- PrA: EOQ = $\sqrt{2 * 28{,}670 * 1{,}000 / 20} \approx 1{,}693$
- PrB: EOQ = $\sqrt{2 * 237{,}370 * 1{,}200 / 20} \approx 5{,}337$
- PrC: EOQ = $\sqrt{2 * 51{,}831 * 1{,}000 / 20} \approx 2{,}277$
- PrD: EOQ = $\sqrt{2 * 13{,}056 * 1{,}200 / 20} \approx 1{,}252$

Total Annual Cost = [(Annual Demand / EOQ) * Order Cost] + [(EOQ / 2) * Holding Cost]

- PrA: Total Annual Cost ≈ [(28,670 / 1,693) * 1,000] + [(1,693 / 2) * 20] ≈ €33,864
- PrB: Total Annual Cost ≈ [(237,370 / 5,337) * 1,200] + [(5,337 / 2) * 20] ≈ €106,786
- PrC: Total Annual Cost ≈ [(51,831 / 2,277) * 1,000] + [(2,277 / 2) * 20] ≈ €45,532
- PrD: Total Annual Cost ≈ [(13,056 / 1,252) * 1,200] + [(1,252 / 2) * 20] ≈ €25,033

Likewise, the EOQ values and other metrics in the table calculated using the formulas and the data for each product.

Table 2. EOQ metrics

|  | **PrA** | **PrB** | **PrC** | **PrD** |
|---|---|---|---|---|
| AnnualDemand | 25,875 | 162,138 | 50,420 | 37,515 |
| EOQ | 1,802 | 5,958 | 1,433 | 3,076 |
| Total AnnualCost | € 32,457 | € 60,288 | € 35,503 | € 56,827 |
| Total AnnualProfit | € 275,837 | € 1,119,663 | € 496,014 | € 1,245,023 |
| Expected Proportion of LostOrder | 0.025 | 0.038 | 0.095 | 0.073 |
| LeadTime Variability | 0.360 | 0.041 | 0.154 | 0.021 |
| StdDev of LeadTimeDemand | 631 | 93 | 206 | 21 |
| SafetyStock | 1,810 | 269 | 952 | 49 |
| Reorder Point (ROP) | 2,515 | 4,160 | 3,218 | 834 |

### 4.1 Risk factors

From Table 1, we assess the risks of holding cost, stock out, and supplier performance for each item to calculate the risk metrics.

- Holding cost risk (HCR): $Holding\ Cost\ per\ Unit = Purchase\ Cost * Holding\ Cost\ Rate$, where the holding cost rate is the holding cost per unit per day.
- Stock out risk (SOR): $Service\ Level = (Demand\ during\ Lead\ Time + Safety\ Stock) / (Starting\ Inventory + Scheduled\ Receipts\ during\ Lead\ Time)$, where safety stock is the amount of inventory held in addition to the expected demand to buffer against uncertainties in demand and lead time.

- Here, we assumed a safety stock of one standard deviation of demand during lead time. The lower the service level, the higher the stock out risk.
- Supplier performance risk (SPR): We considered the lead time and the supplier's probability of meeting the lead time. The probability of meeting the lead time is highest for item PrB (1.00) and lowest for item PrD (0.23). Therefore, the item PrD has the highest SPR.
- $Inventory\ Holding\ Costs\ (IHC) = (Average\ Inventory\ Level) * (Holding\ Cost\ per\ Unit)$
  $Average\ Inventory\ Level = (Order\ Quantity\ /\ 2) + Safety\ Stock$, where the order quantity is the quantity of items ordered each time and the safety stock is the amount of inventory held in addition to the expected demand to buffer against uncertainties in demand and lead time.
- $Expected\ backorder\ costs\ (BOC): = (ExpectedBackorders) * (BackorderCostperUnit)$
  $Expected\ Backorders = (1 - Service\ Level) * (Demand\ during\ Lead\ Time)$, where the service level is the probability of meeting customer demand during the lead time.
- The expected fill rate (EFR): demand that can be fulfilled immediately from available inventory. $EFR = Service\ Level + (1 - Service\ Level) * (Starting\ Inventory\ /\ Demand\ during\ Lead\ Time)$.

Table 3. Risk metrics

| Product | HCR | SOR | IHC | BOC | EFR |
| --- | --- | --- | --- | --- | --- |
| PrA | € 136.80 | 0.984 | € 19,517.88 | € 44.94 | 0.991 |
| PrB | € 7.00 | 0.931 | € 3,167.33 | € 46.23 | 0.995 |
| PrC | € 63.60 | 0.964 | € 28,695.74 | € 188.64 | 0.988 |
| PrD | € 777.00 | 0.881 | € 5,333.59 | € 8,226.20 | 0.943 |

The effectiveness of (p, Q) and (r, Q) policies are investigated in the following section to generate a comparative analysis.

## 5. Empirical investigation

Simulation is used to calculate the appropriate inventory levels in both the (r, Q) and (p, Q) systems.

### 1.1 Periodic review systems (p, Q)

Here, we review our inventory levels at specific intervals (typically at the end of each review period), and we place an order to replenish the stock up to a predetermined quantity known as the "order-up-to point" (OUP). Fig. 3 displays a curve representing the total cost as a function of the order quantity Q, with the optimal order quantity Q* corresponding to the minimum point on the curve (p). The relevance of this plot resides in its capacity to find the best order quantity (Q*) for the (p, Q) inventory policy, which helps minimize costs, maximize profits, and make inventory management more robust against demand and lead time uncertainty.

The algorithm accurately simulates future events by placing an order to restock M units if the date falls within the evaluation window. It estimates inventory and profit behavior under different scenarios, identifying product prices, order costs, and holding costs.

$$Annual\ profit = Revenue - Product\ costs - Ordering\ costs - Holding\ costs$$

where:

- $Revenue = Units\ sold * Selling\ price\ per\ unit$
- $Product\ costs = units\ ordered * unit\ cost\ per\ unit$

- ordering costs = Number of orders placed ∗ Ordering cost per order
- holding costs = Average inventory level ∗ Holding cost per unit per day ∗ Number of days in the year

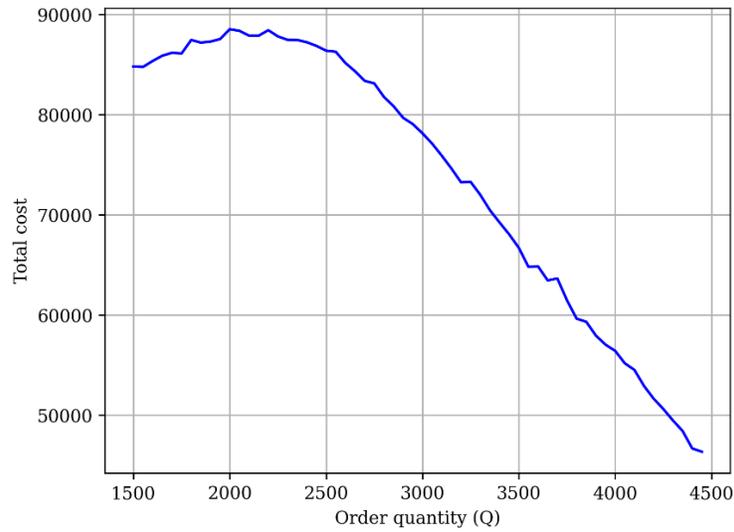

Fig. 3. (p, Q) inventory policy for P=80

Fig. 4 displays the simulation for PrA with the order up to point (OUP) = 2071. We calculate backorders for each day by subtracting the inventory level from the daily demand. If the inventory level > demand, backorders = zero, alternate backorders = excess demand. Fig. 3 displays the plot where the backorders are in red with a dashed line.

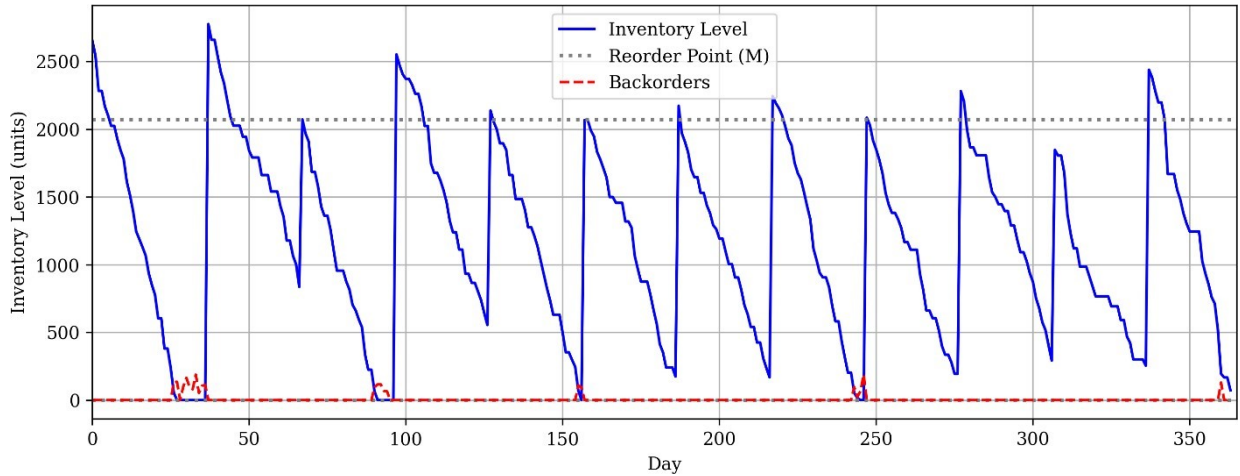

Fig. 4. (p, Q) model - Inventory simulation (PrA), number of orders 12

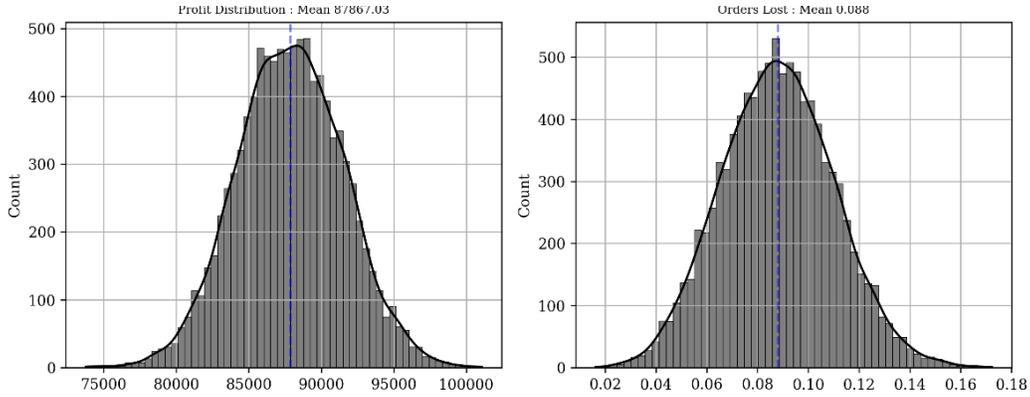

Fig. 5. Distribution plot of profit and lost sales for PrA

To determine the value of order up to point (OUP), we performed simulations of the results for a range of OUP values and calculated the associated profit. A product's optimal OUP may change based on market conditions and production expenses. Fig. 6 shows the computed outcomes for PrA numbers between 1000 and 3000. This resulted in an optimal expected profit of $87,863 (in subsequent simulation runs, $87,992 was optimal). Similar evaluations were carried out on the remaining products, and their optimal values were estimated. Because the profit function is linear in the product's demand and costs, the maxima can be computed.

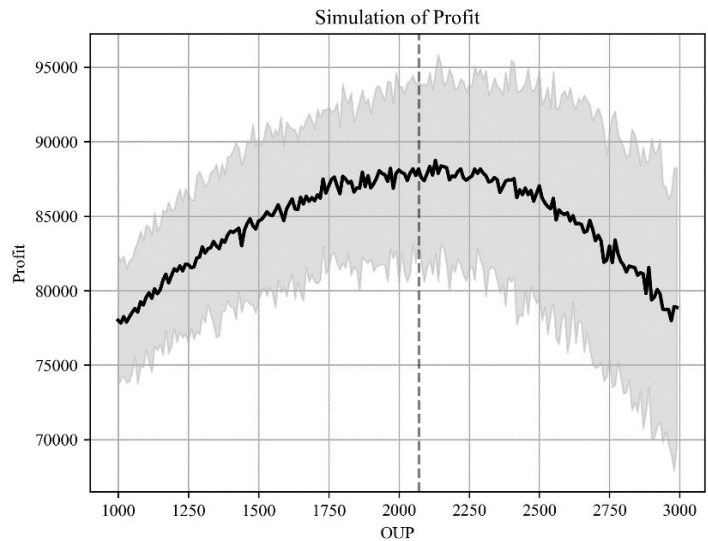

Fig. 6. Simulations output

Using a review time of one month, Table 4 shows the decision variables for each product, including OUP, projected yearly profit, percentage of orders lost over the year, and safety stock.

Table 4. Output of Periodic review

|  | **PrA** | **PrB** | **PrC** | **PrD** | **Total Profit** |
|---|---|---|---|---|---|
| OUP | 2071 | 18,424 | 4154 | 1305 |  |
| Expected Profit | € 88,123.36 | € 372,141.64 | € 166,150.02 | € 320,423.99 | € 944,839.01 |
| Profit std dev | € 5197.05 | € 1400.13 | € 6372.92 | € 29,471.03 | |
| Lost Orders | 0.07 | 0.01 | 0.05 | 0.07 | |
| Safety stock | 61 | 43 | 51 | 5 | |

Table 4 provides useful information for IM decision-making, including how to set order quantities, safety stock levels, and service levels to optimize profit and minimize lost sales. The standard deviation and orders lost provide insights into demand variability and the potential risk of stockouts.

## 1.2 Continuous review (r, Q)

In this policy, we order our products based on a fixed threshold, and as soon as the inventory reaches the threshold, we order a predetermined quantity from the supplier. This threshold is known as a reorder point. So, the order quantity (Q) and the reorder point (r) are critical decision variables in this policy. The optimal values of these variables depend on several factors, such as demand distribution, lead time, and the costs associated with ordering and holding. When an order of Q units comes, the number of things in stock goes up. Depending on the probabilistic demand, the supply goes down at different rates. A new order is placed when the reorder point is met. To account for fluctuations in demand and lead time, safety stock, or excess inventory, is frequently added to the reorder point. This safety stock reduces the possibility of running out of stock before the arrival of a new order. (Anderson et al., 2018). The changes are made according to the algorithm. To determine the optimal order quantity for each product under a (r, Q) policy using the following formula:

$$Q^* = \sqrt{(2DS/H)} \qquad (3)$$

where $Q^*$ is the optimal order quantity, D is the average demand per unit time, S is the fixed order cost, and H is the holding cost per unit per unit time. Using the data from Table 1, the optimal order quantity for each product can be computed as follows:

- $PrA: D = 103.50, S = 1000, H = 20, thus\ Q^* = \sqrt{((2 * 103.50 * 1000)/20)} = 205.94$
- $PrB: D = 648.55, S = 1200, H = 20, Q^* = \sqrt{((2 * 648.55 * 1200)/20)} = 810.329$
- $PrC: D = 201.68, S = 1000, H = 20, Q^* = \sqrt{((2 * 201.68 * 1000)/20)} = 126.768$
- $PrD: D = 150.06, S = 1200, H = 20, Q^* = \sqrt{((2 * 150.06 * 1200)/20)} = 126.768$

These can be adjusted based on business constraints and other considerations, such as storage capacity or minimum order quantities. The program performs a daily stock count and reorder point comparison. If the stock amount is lower than the reorder threshold, an order is submitted. The lead period for the product in question must have passed before the stock can be realized. It uses the same decision reasoning as the (p, Q) algorithm to change the stock. The method used to estimate future profits and losses in order volume is very comparable to that used in the policy under regular review.

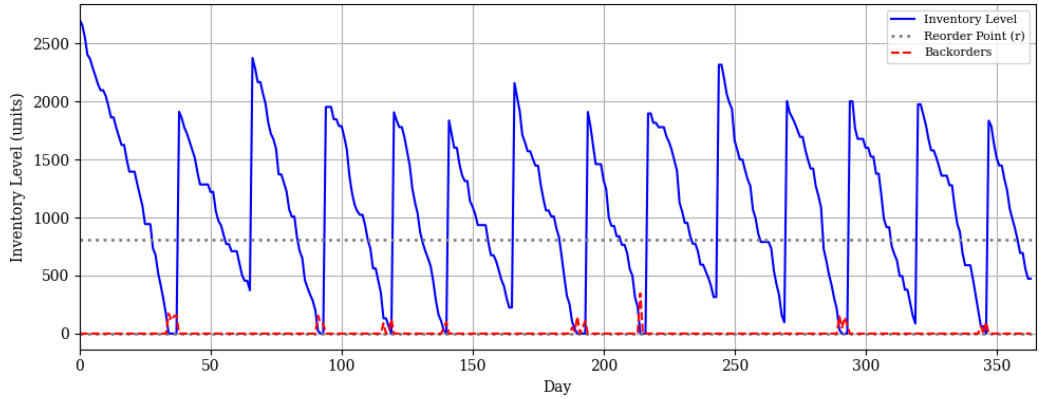

Fig. 7. (r, Q) review (PrA) with stochastic demand ($d \sim N(\mu_d, \sigma_d^2)$): Number of orders 13

Again, 10,000 simulations were run for each product based on these findings. We calculate the average profit, standard deviation, and percentage of cancelled orders using these figures for an order amount for 2002 and a re-order point of 812. Fig. 8 displays the distribution plot for PrA.

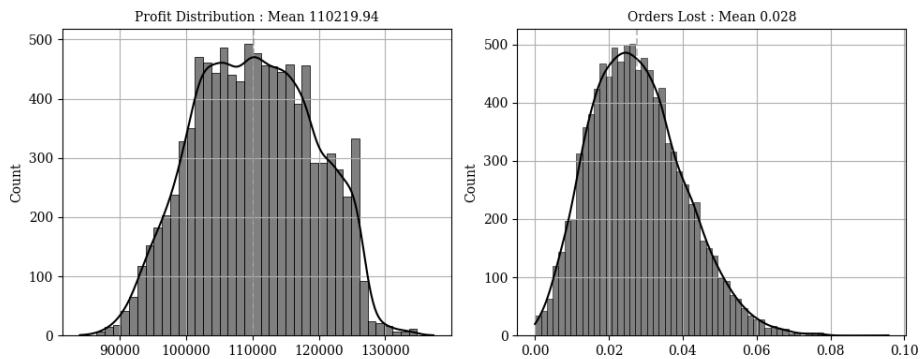

Fig. 8. Distribution plot - Profit & lost orders (PrA) – (r, Q) policy

Table 5 shows the expected annual profits and the proportion of orders lost for each product based on the optimal reorder points and order quantities for this policy.

Table 5. Output of (r, Q)

|  | **PrA** | **PrB** | **PrC** | **PrD** | **Total Profit** |
|---|---|---|---|---|---|
| OUP | 1440 | 22270 | 2570 | 1200 | |
| Reorder Point | 870 | 2790 | 2580 | 1130 | |
| Expected Profit | € 114,126.35 | € 490,888.15 | € 202,588.73 | € 395,753.13 | € 1,203,356.36 |
| Profit std dev | € 6349.42 | € 3251.43 | € 7767.93 | € 3,5397.52 | |
| Lost Orders | 0.03 | 0.06 | 0.02 | 0.02 | |
| Safety stock | 16,623 | 8,421 | 19,979 | 88,882 | |

### 1.3 Risk assessment

A summary of the risks assessed is displayed in Appendix 1 (Table 7).

- Purchase Cost Risk (PCR): The potential cost of purchasing additional inventory, based on the expected purchase cost and the standard deviation of the purchase cost.
- Lead Time Risk (LTR): The potential risk of running out of stock due to delays in the delivery of inventory.
- Quality Risk (QTR): The potential cost of holding or purchasing low-quality inventory, based on the expected profit and the standard deviation of profit.

The metrics and risks are defined as follows:

- Expected Inventory Holding Costs Risk: The risk is low due to low holding cost values, but its impact could be moderate to high if costs increase significantly, resulting in a low to medium overall risk rating.
- Expected Backorder Costs Risk: The risk of backorder costs is low to moderate, with a potential impact on customer dissatisfaction and lost sales, resulting in an overall risk rating of low to medium.
- Expected Fill Rate Metric: A high expected fill rate suggests a low likelihood of stockouts and lost sales, while a low expected fill rate suggests a high likelihood of stockouts and lost sales.
- Sustainability Risk: This risk relates to the potential impact of the item on the environment, social welfare, and ethical considerations. In the table, we assigned a likelihood rating of low and an impact rating of low, indicating that there is a minimal risk of sustainability issues affecting the item.
- Purchase Cost Risk: For OUP and Reorder Point, we assigned a likelihood rating of low and an impact rating of low, indicating that there is a minimal risk of purchase cost fluctuations affecting these items.
- Quality Risk: It is assessed with varying likelihood and impact ratings. For the Expected Profit item, it is rated as low to medium, suggesting a somewhat higher level of risk. However, for OUP and Reorder Point, both likelihood and impact ratings are low, indicating minimal risk regarding quality issues.
- Lead Time Risk: We assigned a likelihood rating of low to medium and an impact rating of low to medium for the Expected Profit item. For OUP and Reorder Point, we assigned a likelihood rating of low and an impact rating of low, indicating that there is a minimal risk of lead time issues affecting these items.

Most risks are classified as "low to medium" risks according to the risk assessments metrics, indicating that they are typically managed.

### 1.4 Comparative analysis

The expected profit for the periodic review (p, Q) is € 944,839.01 and the reorder quantity (r, Q) is € 1,203,356.36, which is approx. 12.73% higher. The (r, Q) approach here requires less safety stock and a lower reorder point. The reduction in inventory held leads to less waste and energy consumption, which makes this more environmentally sustainable than the periodic review approach. The effectiveness of continuous review depends on specific conditions: (1) if the cost of placing an order is high, then the (r, Q) policy is effective since it involves placing smaller, more frequent orders; (2) when the lead time and demand variability co-exist, which is a stochastic nature of the demand pattern. Moreover, our objective is to maximize service level, which is linked to profit maximization and hence (r, Q) is appropriate because it provides more timely information on inventory levels and allows for more immediate action to be taken to replenish inventory when necessary.

To properly optimize continuous review, lead time variability, safety stock, ordering costs, demand variability, and other relevant elements must be considered. By combining observed data and reacting to changing conditions, a data-driven method is employed involving Gaussian Process Regression (GPR) and conditioning functions.

a)   *Gaussian Process Regression (GPR):*

GPR models the objective function $f(x)$ as a Gaussian Process:

$$f(x) \sim GP(m(x), k(x, x')) \qquad (4)$$

here $m(x)$ and $k(x, x'))$ are the mean and covariance functions and denoted as:

$$\left.\begin{array}{l} m(x) = (f(x)) \\ k(x, x') = E\{[f(x) - m(x)][f(x') - m(x')]^T\} \end{array}\right] \qquad (5)$$

The covariance function specifies how the function values change together as the inputs vary. It helps capture patterns, smoothness, and relationships in the data.

b)   *Conditioning Function ($\kappa(x)$):*

The conditioning function introduced here to enhance the adaptability of the optimization algorithm as it accumulates more data. It modifies the mean function $m(x)$ based on observed data or prior knowledge:

$$m_c(x) = m(x) + \kappa(x) \qquad (6)$$

here $m_c(x)$ is the conditioned mean function. The conditioning function defined as:

$$\kappa(x) = \alpha * u(x) \qquad (7)$$

where $\alpha$ = scaling factor and $u(x)$ = a function that captures the influence of observed data on the mean function. The combined objective function is:

$$f_{combined}(x) = m_c(x) + f(x) \qquad (8)$$

The goal is to find the optimal input $x^*$ that maximizes or minimizes the combined objective function ($x^* = arg\ max_x\ f_{combined}(x)$). In practical terms, this means we are looking for the input that gives us the best balance between our prior expectations (captured by $m(x)$) and the observed data (captured by $\kappa(x)$), while optimizing the original objective function $f(x)$.

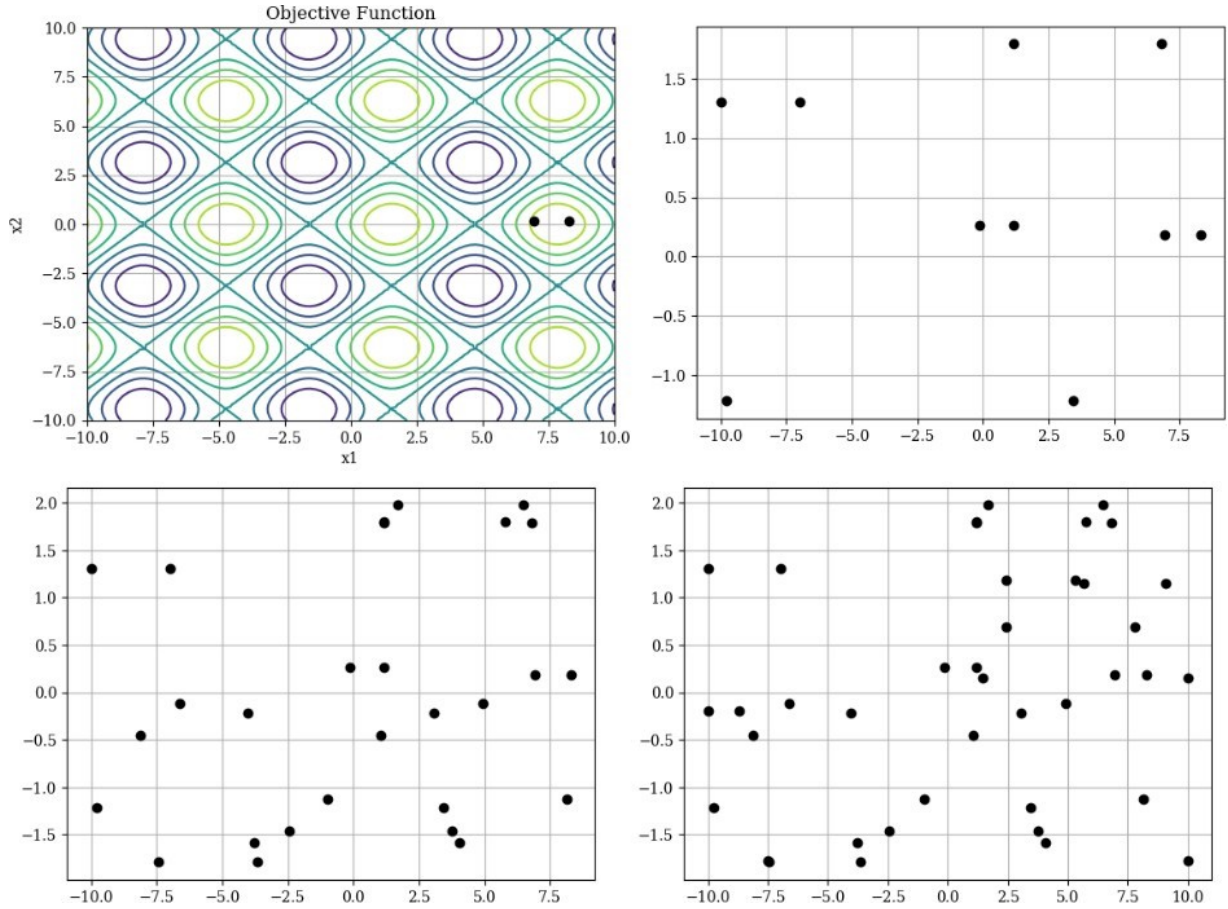

Fig 9.  Optimization progression (truncated plots)

Fig. 9 displays the progression of the BO algorithm over multiple iterations. The black dots represent the points sampled by the algorithm, and the contour lines represent the objective function's contours. As the algorithm progresses, it tries to sample points in areas with higher objective function values, indicated by the contour lines getting closer together.

The algorithm should converge to a global optimum or a local optimum, depending on the nature of the objective function. Fig. 10 displays the GPR with the EI acquisition function at each iteration of the BO loop. The colored points represent the observation points, where the color represents the value of the objective function at that point. The black contours represent the predicted values of the objective function by the GPR model. The black dots represent the next best point to sample based on the EI acquisition function. From the plots, we can infer how the GPR model is gradually improving its predictions of the objective function with each iteration and how the acquisition function is guiding the search towards the areas of the search space that are most likely to contain the global optimum.

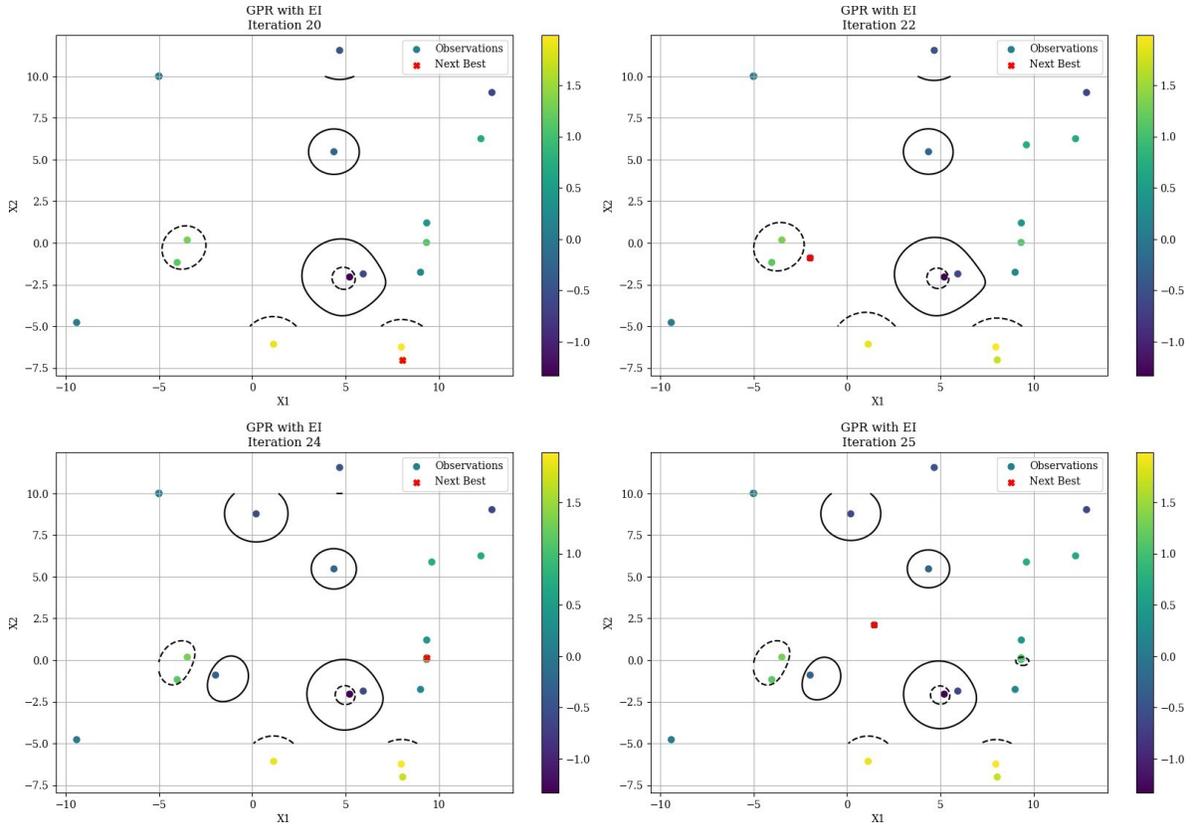

Fig. 10. GPR with the acquisition function and the observation points

We also observe how the black dots are moving towards the areas where the GPR model is uncertain about the objective function values, which is where the EI acquisition function is most likely to suggest sampling. Finally, we can see how the observation points are gradually becoming more concentrated around the global optimum as the optimization loop progresses.

Further, the optimization algorithm was tuned using a conditioning multivariate Gaussian distribution, which involves updating the prior distribution based on the observed data. The objective here is to obtain a flexible and powerful approach for optimizing complex functions. This allowed us to incorporate the uncertainty in the observations and adjust the model accordingly for better predictions of the optimal solution.

Fig. 11 displays the conditioning of the multivariate Gaussian distribution means updating the distribution based on observed data. The target function plot shows the original target function that we are trying to optimize, which is a two-dimensional Gaussian distribution with mean [1,2] and covariance matrix $\begin{bmatrix} 2 & 0.5 \\ 0.5 & 1 \end{bmatrix}$. The plot shows that the peak of the distribution is located near the mean of [1, 2]. The goal of this plot is to illustrate how Bayesian optimization can be used to find the maxima of the conditioned distribution. In the right plot, Bayesian optimization aims to find the peak, which may not necessarily align with the peak of the original target distribution.

Since it models the objective function as a multivariate Gaussian distribution, it can capture correlations between input parameters and incorporate prior knowledge about the objective function. We used a combination of GPR and conditioning functions to improve optimization performance by leveraging the strengths of both approaches. The GPR model provides a probabilistic estimate of the unknown function, while the conditioning function biases the search towards promising regions. The optimization process can be made more efficient and effective by selecting appropriate mean and covariance functions and designing a suitable conditioning function based on prior knowledge or intuition.

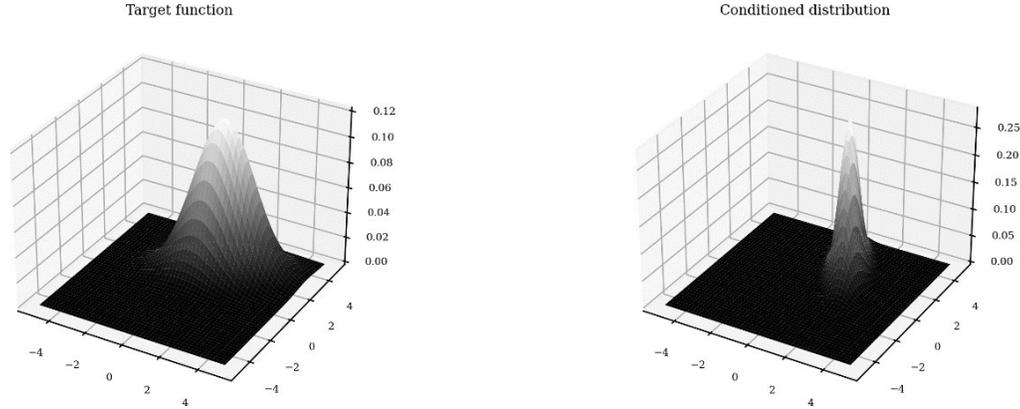

Fig. 11. Conditioning multivariate Gaussian

The combined objective function was optimized using an acquisition function, expected improvement (EI), probability of improvement (PI), to select the next input point to evaluate. The EI function defined as

$$EI(x) = \int_{-\infty}^{f_{min}} (f_{min} - f_{combined}(x))^+ p(f_{combined}(x)|D) df_{combined}(x) \quad (9)$$

Here, $f_{min}$ is the minimum observed value of the objective function, $p(f_{combined}(x)|D)$ is the posterior probability distribution of $f_{combined}(x)$ given the observed data D, and $(\cdot)^+$ denotes the positive part of the function.

Fig. 12 visually demonstrates how the GP regression model evolves as it observes more data points and refines its estimate of the unknown function. The purpose of this plot is to visualize how the GP model approximates the unknown function, considering both the observed data points and the conditioning function.

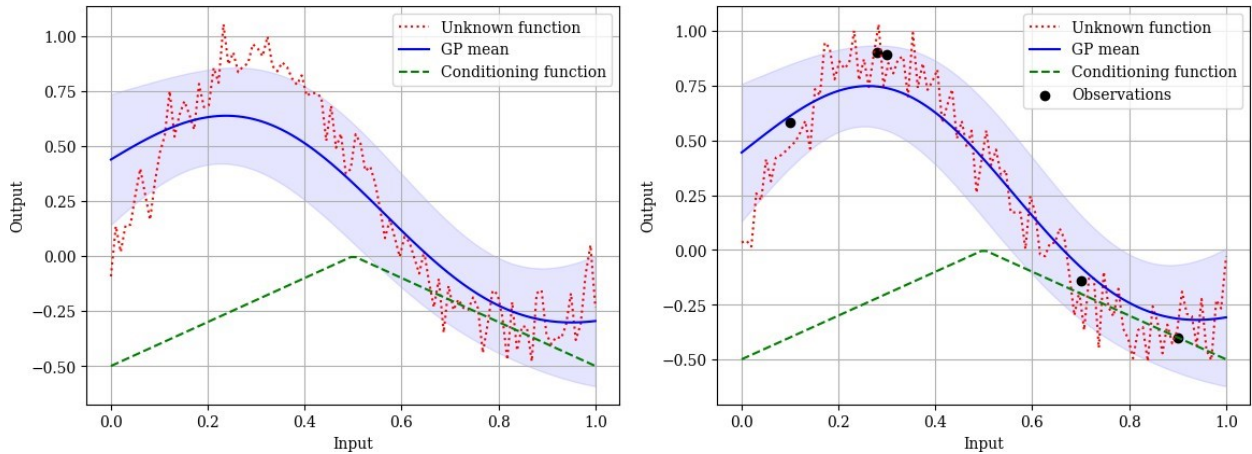

Fig. 12. GPR with & without conditioning function

The shaded region in Fig. 12 from left hand side plot is around the mean with the 95% confidence interval, which is wider in regions where there are fewer observed data points. The Gaussian Process Regression without conditioning on $x_0$ captures the general trend of the data, but there is significant uncertainty in the predictions in some regions. The plot on the right-hand side can use the information from the observed data points with the same $x_0$ value to make more accurate predictions. It can be inferred that conditioning on $x_0$ can significantly improve the accuracy of the Gaussian process regression in regions where there are observed data points with the same $x_0$ value.

To this end, this empirical investigation explored different inventory management strategies by conducting simulations to estimate optimal inventory policies, assess risks, and introduce Bayesian optimization (BO) with Gaussian Process Regression (GPR) and conditioning functions. The key findings suggest the (r, Q) continuous review system outperforms the (p, Q) periodic review system in expected profits and inventory efficiency. BO, coupled with GPR and conditioning functions, is presented as a powerful approach for optimizing complex functions in inventory management, considering uncertainty, and adapting to changing conditions.

The best order quantities (annual): Best Solution: [584, 905, 3793, 1214] and best value (Profit): 27736262 considering the bounds = [(0, 5000), (0, 5000), (0, 5000), (0, 5000)].

c)   Practical application

To this end, the study provides a comprehensive analysis of inventory management strategies, including periodic review (p, Q) and continuous review (r, Q) systems, and introduces Bayesian optimization with Gaussian Process Regression (GPR) and conditioning functions as a promising approach for optimizing inventory policies under uncertainty. The potential real-life applications can be found in retail supply chains, manufacturing, pharmaceuticals, e-commerce fulfilment centres, and food and beverage. Benefits for industry managers include cost savings, improved profit margins, enhanced customer service, reduced stockout risk, environmental sustainability, adaptability to changing conditions, data-driven decision-making, risk mitigation, and competitive advantage. Retailers can optimize inventory levels to minimize holding costs and stockout risks, while manufacturing can optimize raw material and component inventory management for industries with fluctuating demand and lead times. Pharmaceutical companies can manage inventory of critical drugs and medical supplies, while e-commerce businesses can optimize fulfilment centre operations.

In this use case, with the stochastic demand, the continuous review system reduces waste and energy consumption, making inventory management more environmentally sustainable. Integration of Bayesian optimization with Gaussian Process regression and conditioning functions provides a dynamic approach to inventory management, allowing stakeholders to adapt their strategies based on evolving market conditions and demand patterns. Data-driven decision-making is essential, and risk assessment metrics can help identify potential risks related to purchase costs, lead times, quality, and more. Implementing advanced inventory management strategies can provide a competitive edge in the market.

## 1.5 Sensitivity analysis

Table 6 displays the sensitivity analysis by varying each variable one at a time while keeping the others constant. This helps to understand how changes in each variable affect the overall results. These are estimated numbers assuming that the relationships between the variables are linear.

Table 6. Sensitivity analysis by varying each variable one at a time

| Variable | Change | PrA | PrB | PrC | PrD |
|---|---|---|---|---|---|
| OUP | 10% increase | € 1,323,834.40 | € 25,273.84 | € 219,500.60 | € 429,894.00 |
| | 5% decrease | € 1,046,974.30 | € 20,266.16 | € 185,676.85 | € 362,010.48 |
| Reorder Point | 10% increase | € 1,130,191.28 | € 23,710.08 | € 195,814.78 | € 382,408.96 |
| | 5% decrease | € 1,122,545.42 | € 21,830.24 | € 189,787.13 | € 371,250.88 |
| Expected Profit | 10% increase | € 1,343,247.95 | € 26,256.21 | € 225,308.46 | € 439,710.77 |
| | 5% decrease | € 962,004.94 | € 18,423.09 | € 180,869.00 | € 354,791.46 |
| Profit Std Dev | 10% increase | € 1,305,506.25 | € 25,739.81 | € 219,789.29 | € 428,368.53 |
| | 5% decrease | € 1,195,748.85 | € 22,743.56 | € 199,386.23 | € 388,788.49 |
| Lost Orders | 10% increase | € 1,199,999.88 | € 24,079.94 | € 200,039.78 | € 390,119.90 |
| | 5% decrease | € 1,200,003.78 | € 23,999.94 | € 200,019.89 | € 390,039.92 |
| Safety Stock | 10% increase | € 1,373,372.16 | € 27,526.32 | € 232,141.80 | € 454,631.04 |
| | 5% decrease | € 1,291,179.52 | € 25,869.84 | € 218,057.31 | € 426,564.80 |

Taking every variable vary independently of each other and assuming a 10% increase in PrA, PrB, PrC, and PrD and a 5% decrease in PrA, PrB, PrC, and PrD, the sensitivity analysis table displayed at Appendix 2 (Table 8 & 9).

**Conclusion**

This study examined two inventory management strategies: the (r, Q) continuous review system and the (p, Q) periodic review system. It focused on optimizing inventory policies, risk assessment, and risk management. The continuous review system outperforms the periodic review system in terms of expected profits and inventory efficiency. Bayesian optimization, combined with Gaussian Process Regression and conditioning functions, is introduced as a powerful approach for optimizing complex inventory functions. The study also considered sustainability risk and the contribution of conditioning functions. Bayesian optimization is found to be a robust method for optimizing inventory policies in uncertain and complex situations. The study emphasized the importance of a combination of advanced optimization techniques, thorough risk assessment, and environmental considerations in inventory management.

**<<<Computer code and data are available on request>>>**

**Appendix 1**

Table 7. Potential risk factors

| Risk Factor | Description | Potential Impact | Potential Ethical Implications | PrA | PrB | PrC | PrD |
|---|---|---|---|---|---|---|---|
| Stockout Risk | The risk of running out of inventory and not being able to meet customer demand. | Loss of revenue, customer dissatisfaction, reputational damage. | Failure to meet customer expectations, unfair treatment of customers, lack of transparency. | 0.984 | 0.931 | 0.964 | 0.881 |
| Holding Cost Risk | The risk of incurring high inventory holding costs due to excess inventory. | Financial losses, inefficient use of resources. | Wastefulness, failure to optimize resources, environmental impact. | € 136.80 | € 7.00 | € 63.60 | € 777.00 |
| Purchase Cost Risk | The risk of incurring high purchase costs due to fluctuations in prices. | Financial losses, reduced profit margins. | Lack of supplier diversity, reliance on unethical suppliers, conflict of interest. | € 44.94 | € 46.23 | € 188.64 | € 8,226.20 |
| Lead Time Risk | The risk of not receiving inventory on time due to delays or variability in lead time. | Stockouts, customer dissatisfaction, reduced productivity. | Failure to plan, lack of supplier diversity, lack of transparency. | 0.991 | 0.995 | 0.988 | 0.943 |
| Quality Risk | The risk of receiving inventory that does not meet quality standards. | Customer dissatisfaction, product recalls, reputational damage. | Lack of supplier due diligence, failure to prioritize quality, lack of transparency. | 0.05 | 0.035 | 0.036 | 0.031 |
| Sustainability Risk | The risk of negative environmental or social impact due to inventory management practices. | Environmental damage, human rights violations, reputational damage. | Failure to prioritize sustainability, lack of transparency, unethical sourcing practices. | Low | Low | High | Medium |

# Appendix 2

Table 8. Risk Estimate from (r, Q) model.

| Item | HCR | SOR | SPR | IHC | BCR | EFR | PCR | QTR | LTR | STR | Overall rating |
|---|---|---|---|---|---|---|---|---|---|---|---|
| OUP | Low | Moderate | Low | Low | Low | High | Low | Low | Low | Low | Low to Medium |
| Reorder Point | Low | Moderate | Low | Low | Low | High | Low | Low | Low | Low | Low to Medium |
| Expected Profit | Low | High | Low | Low to Medium | Low to Medium | High | Low to Medium | Low to Medium | Low | Low | Medium to High |
| Lost Orders | Low | Moderate | Low | Low | Low | High | Low | Low | Low | Low | Low to Medium |

Table 9. Sensitivity analysis (r, Q) policy with 10% and 5% variation in data

|  | PrA (+10%) | PrA (-5%) | PrB (+10%) | PrB (-5%) | PrC (+10%) | PrC (-5%) | PrD (+10%) | PrD (-5%) |
|---|---|---|---|---|---|---|---|---|
| OUP | 1,239,690 | 1,081,791 | 24,497 | 20,958 | 2,827 | 2,473 | 1,314 | 1,146 |
| Reorder Point | 957 | 828 | 2,790 | 2,402 | 2,580 | 2,248 | 1,130 | 978 |
| Exp. Profit | € 478,972.97 | € 75,280.50 | € 539,976.97 | € 362,079.79 | € 222,847.60 | € 182,327.90 | € 435,328.44 | € 356,177.46 |
| Profit std dev | € 6,984.36 | € 5,914.52 | € 3,576.58 | € 3,066.22 | € 8,544.72 | € 7,587.05 | € 38,791.27 | € 34,000.64 |
| Lost Orders | 0.033 | 0.028 | 0.066 | 0.057 | 0.022 | 0.019 | 0.022 | 0.019 |
| Safety stock | 18,286 | 14,978 | 9,263 | 7,560 | 21,977 | 18,056 | 97,770 | 80,134 |